\documentclass[twocolumn,aps,ams,showpacs,superscriptaddress]{revtex4}

\usepackage{graphicx,color,soul}
\usepackage{latexsym}
\usepackage{amsmath,amssymb}        
\usepackage[draft=false]{hyperref}
\usepackage{mathrsfs}

\usepackage{mathrsfs}

\begin{document}


\title{Rotation curves of ultralight BEC dark matter halos with rotation}


\author{F. S. Guzm\'an}
\affiliation{Instituto de F\'{\i}sica y Matem\'{a}ticas, Universidad
              Michoacana de San Nicol\'as de Hidalgo. Edificio C-3, Cd.
              Universitaria, 58040 Morelia, Michoac\'{a}n,
              M\'{e}xico.}

\author{F. D. Lora-Clavijo}
\affiliation{Instituto de Astronom\'{\i}a, Universidad Nacional Aut\'{o}noma de M\'{e}xico,
        AP 70-264, Distrito Federal 04510, M\'{e}xico.}


\date{\today}


\begin{abstract}
We study the rotation curves of ultralight BEC dark matter halos. These halos are long lived solutions of initially rotating BEC fluctuations. In order to study the implications of the rotation characterizing these long-lived configurations we consider the particular case of a boson mass $m=10^{-23}\mathrm{eV/c}^2$ and no self-interaction. We find that these halos successfully fit samples of rotation curves (RCs) of LSB galaxies.
\end{abstract}




\maketitle


\section{Introduction}

The idea of an ultralight spinless particle as dark matter originates at cosmic scales, when it was discovered that the predicted mass power spectrum of fluctuations made of a free real scalar field, minimally coupled to General Relativity in an affective theory, with mass $m_{\phi}\sim 10^{-22} - 10^{-23}eV$ could ameliorate the problem of small structures associated to the standard cold dark matter model \cite{MatosUrena2000}. An interesting property of interpreting this scalar field as a spinless boson is that its condensation temperature is $T_c \sim 1/m^{5/3} \sim \mathrm{TeV}$ for $m_{\phi}\sim 10^{-22}$. This interpretation permits the formulation of the structure formation problem as an evolution problem ruled by the evolution of the condensate, namely the Gross-Pitaevskii equation \cite{GPE} that describes the evolution of the condensate \cite{Urena2014}, with a potential due to the gravity of the fluctuation itself. It was until very recently that the dynamic structure formation problem  was analyzed for such a condensate for the case of an ultralight boson with no self-interaction \cite{Schive}.

Aside of the different emerging questions on the analysis of structure formation within the model, the problem at local scale remains. This problem involves the analysis of the collapse, formation, evolution and virialization of BEC halos that would be ruled by the evolution of a BEC trapped by its own gravitational field, that is, the Gross-Pitaevskii equation coupled to Poisson equation, the GPP system. In fact this local problem has been studied from different angles and some advances have been achieved, which are summarized briefly as follows. Historically, the first time that a solution to the GPP system in spherical symmetry was presented, it was in one of the original papers defining Boson Stars, because it happens that the low-energy and Newtonian limit of the Einstein-Klein-Gordon for a complex scalar field is exactly the GPP system of equations, and numerical solutions of stationary equilibrium configurations were constructed \cite{Ruffini}. Later on, other equilibrium configurations -dubbed gravitational atoms- were constructed, which were solutions to the same GPP system, however for wave functions with nodes; the fact that the solution wave functions had nodes invited to call them excited state solutions as an analogy to the solution of the Schr\"odinger equation for the Hydrogen atom \cite{SinJin,Gleiser1998}. The excited state solutions were in fact applied as models of galactic dark matter, because the RCs of the resulting configurations had very similar properties of observed galactic RCs \cite{SinJin}. These excited states were shown to be unstable though \cite{GuzmanUrena2004,GuzmanUrena2006}. At the same time, other flavors or approaches were developed, for instance another version of the model using the GPP system  to explain the RCs was dubbed the model of quintessential halos \cite{Arbey1,Arbey2}; another example is the fuzzy dark matter model, that is actually a 1D version (not even spherically symmetric in 3D) of the analysis of the collapse of fluctuations obeying the GPP system \cite{fuzzy}. A very complete analysis of BEC spherical solutions can be found in \cite{Chavanis11032050, Chavanis11032054}

Some other BEC dark matter halo models include those constructed in the Thomas-Fermi limit (TF), in which the self-interaction among bosons dominates over the kinetic term in Schr\"odinger equation  \cite{BohemerHarko2007}. Finite temperature effects have also been considered in this regime \cite{FTHarko}, and it has been shown to be a possible solution of the galactic cusp-core problem \cite{BECCuspCoreHarko}. The life-time of these halos has been estimated \cite{HarkoUnstable}, their energy contents \cite{Souza} and 
the condensation process has also been studied \cite{Harko2011}. Recently, the full collapse process of BEC dark matter fluctuations obeying the GPP system has been analyzed in \cite{Chavanis1103.2698,Harko2014}. When compared with observations, galactic halos constructed in the TF limit are capable of fitting RCs, although some uncertainty has been mentioned with respect to the possibility of their formation \cite{HarkoUnstable}.

In this paper we do not consider the model in the TF limit. Instead we consider the system is not dominated by the self-interaction nor the kinetic contribution and we assume the whole GPP system to hold. As mentioned above, equilibrium configurations can be constructed, however some steps forward have been done in more dynamically general situations. For instance, it has been shown that initial fluctuations of the density of the condensate quickly virialize after the turnaround point, in which the cosmic expansion ceases to dominate over the self-gravity of the structure \cite{GuzmanUrena2003}, 
through the gravitational cooling \cite{SeidelSuen1990,GuzmanUrena2006}. It was also shown that in general the evolution of an initial spherical fluctuation ends up approaching one of the ground state equilibrium solutions \cite{GuzmanUrena2006} and also that non-spherical initial fluctuations approach these equilibrium solutions as well \cite{BernalGuzman2006}. Thus equilibrium configurations are stable under a variety of perturbations, including non-spherical perturbations, and also show an attractor type of behavior. These properties together are the reason to consider  ground state solutions a very appealing candidate to be dark matter halos.

In this general scenario the dynamical properties of the halo formation and virialization has therefore been explored, however, unlike the models in the TF limit, RCs have not been explored for already relaxed configurations, and the goal of the present paper is precisely to show their potential at fitting RCs. At this point two alternatives have been explored so far. On the one hand, even though excited states are unstable, the superposition of ground and excited states turned out to be stable and show appealing galactic RCs \cite{BernalUrena2012}. On the other hand, some rotation has been added to ground state configurations that disperse away the density of the condensate and show also appealing RCs \cite{Guzmanetal2014}. The possibility that BEC dark matter halos may have rotation has been pointed out before in \cite{RindlerShapiro2012}, where spheroid and ellipsoid analytic solutions to the GPP system with rotation are studied as rotating BEC dark matter halos in various scenarios, and the results particularly focus on the possibility of vortex formation, but little is said about the comparison with observations, say RCs.

What we do in this paper is to try to answer the question of whether the long-lived rotating or spherical  configurations in \cite{Guzmanetal2014} are capable of fitting observed RCs. For this we focus on a sample of dark matter dominated LSB galaxies, for which, as a first approximation, luminous matter would not contribute significantly to the total mass of the system and thus behave as test particles. As a workhorse we also consider the BEC without self-interaction, in order to be consistent with the only structure formation analysis so far with BEC dark matter at cosmological scale and use the same ultralight boson mass $10^{-23}\mathrm{eV/c}^2$  \cite{Schive}.

The paper is organized as follows, in section \ref{sec:equations} we present the conventions and numerical methods used to solve the GPP system and the methods used to calculate the RCs. In section \ref{sec:RCs} we fit actual RCs and in \ref{sec:conclusions} we discuss the results.

\section{The GPP system}
\label{sec:equations}

\subsection{Equations and numerical scaling}

{\it Units and scaling.} The Gross-Pitaevskii-Poisson (GPP) system of equations describing the evolution of a Bose condensate is

\begin{eqnarray}
i\hbar \frac{\partial \tilde{\Psi}}{\partial \tilde{t}} &=& -\frac{\hbar^2}{2m}\tilde{\nabla}^2 \tilde{\Psi} + \tilde{V}\tilde{\Psi} +\frac{2\pi \hbar^2 \tilde{a}}{m^2} |\tilde{\Psi}|^2\tilde{\Psi}, \nonumber\\
\tilde{\nabla}^2 \tilde{V} &=& 4 \pi G m |\tilde{\Psi}|^2, \label{eq:SPcompleta}
\end{eqnarray}

\noindent where in general the wave function depends on space and time $\tilde{\Psi} = \tilde{\Psi}(\tilde{t},\tilde{{\bf x}})$, $m$ is the mass of the boson, $\tilde{V}$ is the gravitational potential acting as the condensate trap, $\tilde{a}$ is the scattering length of the bosons. This is a coupled system of an evolution equation for $\Psi$ with a potential that is solution of Poisson equation sourced by $|\Psi|^2$.

Before integrating (\ref{eq:SPcompleta}) it is important to remove the constants using the following change of variables $\hat{\Psi} = \frac{\sqrt{4\pi G}\hbar}{mc^2}\tilde{\Psi}$,
$\hat{x} = \frac{mc}{\hbar}\tilde{x}$,
$\hat{y} = \frac{mc}{\hbar}\tilde{y}$,
$\hat{z} = \frac{mc}{\hbar}\tilde{z}$,
$\hat{t} = \frac{mc^2}{\hbar}\tilde{t}$,
$\hat{V} = \frac{\tilde{V}}{mc^2}$,
$\hat{a} \rightarrow \frac{c^2}{2mG}\tilde{a}$, 
so that the numerical coefficients $\hbar,~\hbar^2/m,~2\pi \hbar^2/m^2,~4\pi Gm$ do not appear in (\ref{eq:SPcompleta}). Thus by fixing the boson mass $m$ the hatted variables are fixed.

Going further, the system (\ref{eq:SPcompleta}) is invariant under the transformation $t = \lambda^2 \hat{t}$, 
$x = \lambda\hat{x}$, 
$y = \lambda\hat{y}$, 
$z = \lambda\hat{z}$, 
$ \Psi = \hat{\Psi}/\lambda^2$,
$V = \hat{V} / \lambda^2$,
$a = \lambda^2 \hat{a}$, 
 for an arbitrary value of the parameter $\lambda$ \cite{GuzmanUrena2004}. This rescaling reduces the original system (\ref{eq:SPcompleta}) to the following one
  
\begin{eqnarray}
i \frac{\partial \Psi}{\partial t} &=& -\frac{1}{2}\nabla^2 \Psi + V\Psi +a|\Psi|^2\Psi \label{eq:SchroNoUnits}\\
\nabla^2 V &=& |\Psi|^2, \label{eq:PNoUnits}
\end{eqnarray}

\noindent which is the one we solve numerically. The consequences of this invariance are extremely relevant for our analysis here. The reason is that if one calculates one solution for the non-hatted variables, say for $\lambda=1$, other solutions are automatically found for any other value of $\lambda$, that is, a new solution for the hatted and tilde variables is found just by using a new value of $\lambda$.

{\it Evolution.} The solution is calculated considering Schr\"odinger equation as an evolution equation for $\Psi$, and Poisson equation as a constraint that has to be solved every time it is required during the integration of Schr\"odinger equation. 

We approximate the GPP system (\ref{eq:SchroNoUnits}-\ref{eq:PNoUnits}) using finite differences on a uniformly discretized grid on a spatial  domain described with cartesian coordinates. We solve the system on the spatial domain  $[x_{min},x_{max}]\times [y_{min},y_{max}]\times [z_{min},z_{max}]$ using the 3D Fixed Mesh Refinement (FMR) code developed by us and tested in  \cite{Guzmanetal2014}. The evolution uses a method of lines with a Runge-Kutta integrator. On the other hand, Poisson equation is solved reducing the problem to a 2D slice on the plane $x+y=0$ because we only deal with axial configurations in this work; then the solution is interpolated back into the 3D mesh. The algorithm used to solve Poisson equation on the 2D slice is a successive overelaxation method with optimal acceleration parameter.

\subsection{Rotation curves}

In order to construct the rotation curve of one of our BEC halos, we place various detectors at which we calculate the tangential velocity $v$ of a test particle. The detectors are located at a set of points along the $x$-axis. We assume the test particle describes a circular orbit which implies that the tangent velocity of the test particle is $v(r) = \sqrt{2GM(r)/r}$, where $r$ is the distance from the coordinate center to the detector, and $M(r)$ is the mass contained within a sphere of radius $r$, and calculated as a volume integral in the 3D domain as $M = \int |\Psi|^2 dx dy dz$. Explicitly, if a particular detector is located at $(x,y,z)=(x_d,0,0)$ we calculate $v(x_d)$ as

\begin{equation}
v^2(x_d) = \frac{2 G}{|x_d|} \int |\Psi|^2 dxdydz, \label{eq:rc}
\end{equation}

\noindent where the volume integral is calculated within a sphere of radius $x_d$ on our 3D cartesian grid.

\subsection{Initial data}

Our BEC halos are the relaxed long-lived configurations resulting from the evolution of an initial fluctuation we call protohalo.

The initial data we choose for a proto-halo, is a ground state equilibrium configuration added with an initial angular momentum. When the added angular momentum is zero, the protohalo is actually a ground state equilibrium configuration stable and long-lived since the begining. When some angular momentum is added initially to an equilibrium configuration, it is not a ground state equilibrium configuration anymore, and it takes a time for the configuration to relax and become a long-lived structure that plays the role of one of our BEC halos. Thus the construction of initial data requires a description of the equilibrium configurations. 

{\it Equilibrium configurations.} These are built assuming the wave function is spherically symmetric, a reason to use spherical coordinates, and it is also assumed it is harmonically time dependent $\Psi=\Psi(r,t)=e^{i\omega t}\psi(r)$, which immediately guarantees that $|\Psi|^2$ is time-independent and therefore the potential $V=V(r)$ as well. Under these conditions the GPP system (\ref{eq:PNoUnits}) reduces to 

\begin{eqnarray}
\frac{d^2\psi}{dr^2} + \frac{2}{r} \frac{d\psi}{dr} &=& 2(V+a|\psi|^2 + \omega)\psi, \nonumber\\
\frac{d^2 V}{dr^2} + \frac{2}{r} \frac{d V}{dr}  &=& 4\pi |\psi|^2,\label{eq:eigenvalue}
\end{eqnarray}

\noindent which is an eigenvalue problem for $\psi(r)$ with the boundary conditions of wave function smoothness at the origin and isolation at infinity, that is $\psi(r\rightarrow \infty)\rightarrow0$. This problem is then solved numerically in a finite spatial domain. The boundary condition for the gravitational potential is $V=-M/r$ for $r\rightarrow \infty$ where $M=\int |\psi(r)|^2 r^2 dr$ as described in \cite{GuzmanUrena2004}. The solution is such that given a value for $\psi(0)$ there is a unique eigenvalue $\omega$ satisfying the boundary conditions and plays the role of the eigen-energy of the system. 

The solutions of the eigenvalue problem (\ref{eq:eigenvalue}) can have nodes or not. Those configurations with nodes are called excited state solutions whereas those without nodes are called ground state solutions. In \cite{GuzmanUrena2006} was shown in detail, using numerical simulations, that excited state configurations are unstable and in fact decay into one of the equilibrium configurations through a relaxation process called gravitational cooling that consists in the ejection of density of probability to infinity \cite{SeidelSuen1990,GuzmanUrena2006}.

With this in mind we are only interested in ground state configurations. We then remark that there is a configuration for each different value of $\psi(0)$ that in the end has a specific associated value of $\omega$. Two comments are in turn: i) given the rescaling property of the GPP system, it is not necessary to construct all the possible wave functions and density profiles for each value of $\psi(0)$, because given one, say $\psi(0)=1$, one can construct all the other possible ground state configurations  using simply different values of the scale parameter $\lambda$; ii) since we are interested in the density of the BEC, that is $|\Psi|^2$, $\omega$ does not play a relevant role in our analysis and in fact -if required- the value of this frequency is associated to a specific central value of the wave function $\psi(0)$.

\begin{figure}
\includegraphics[width=7cm]{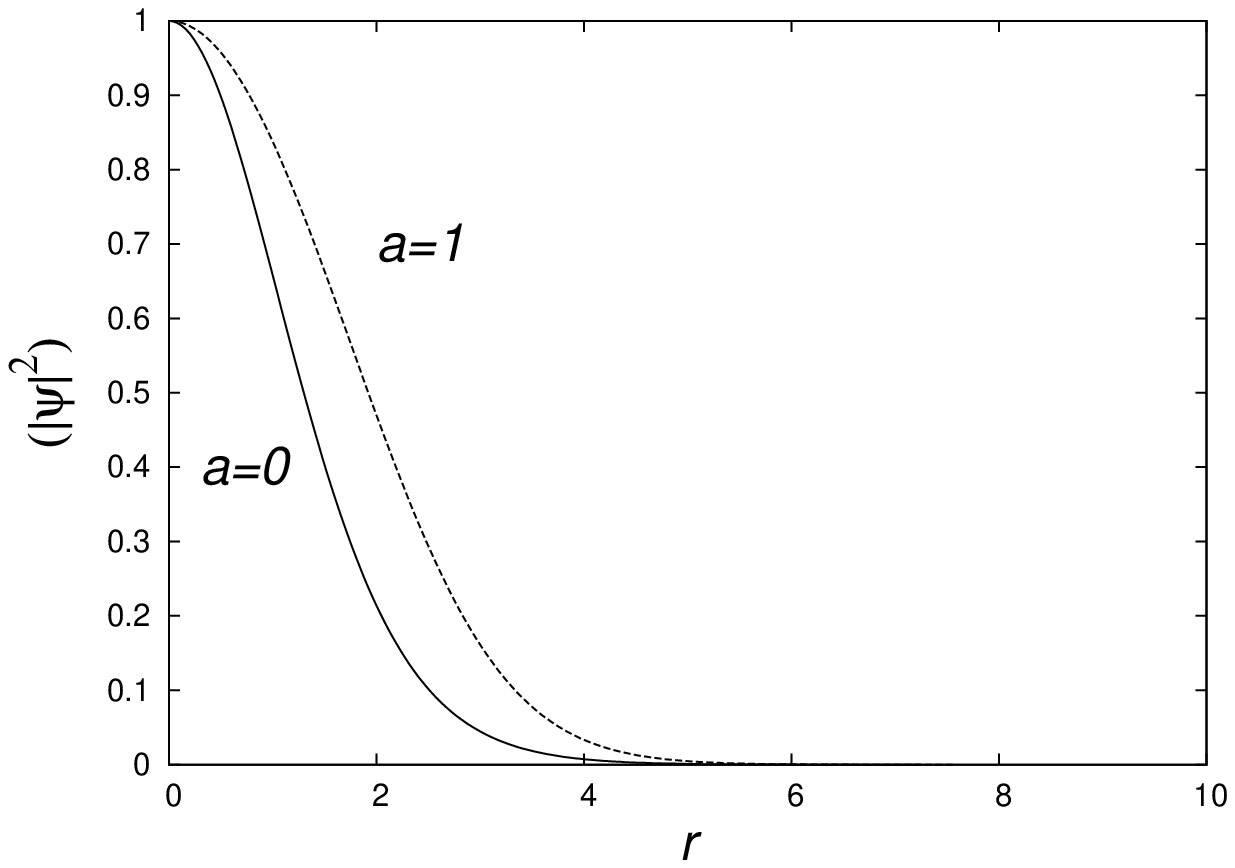}
\includegraphics[width=7cm]{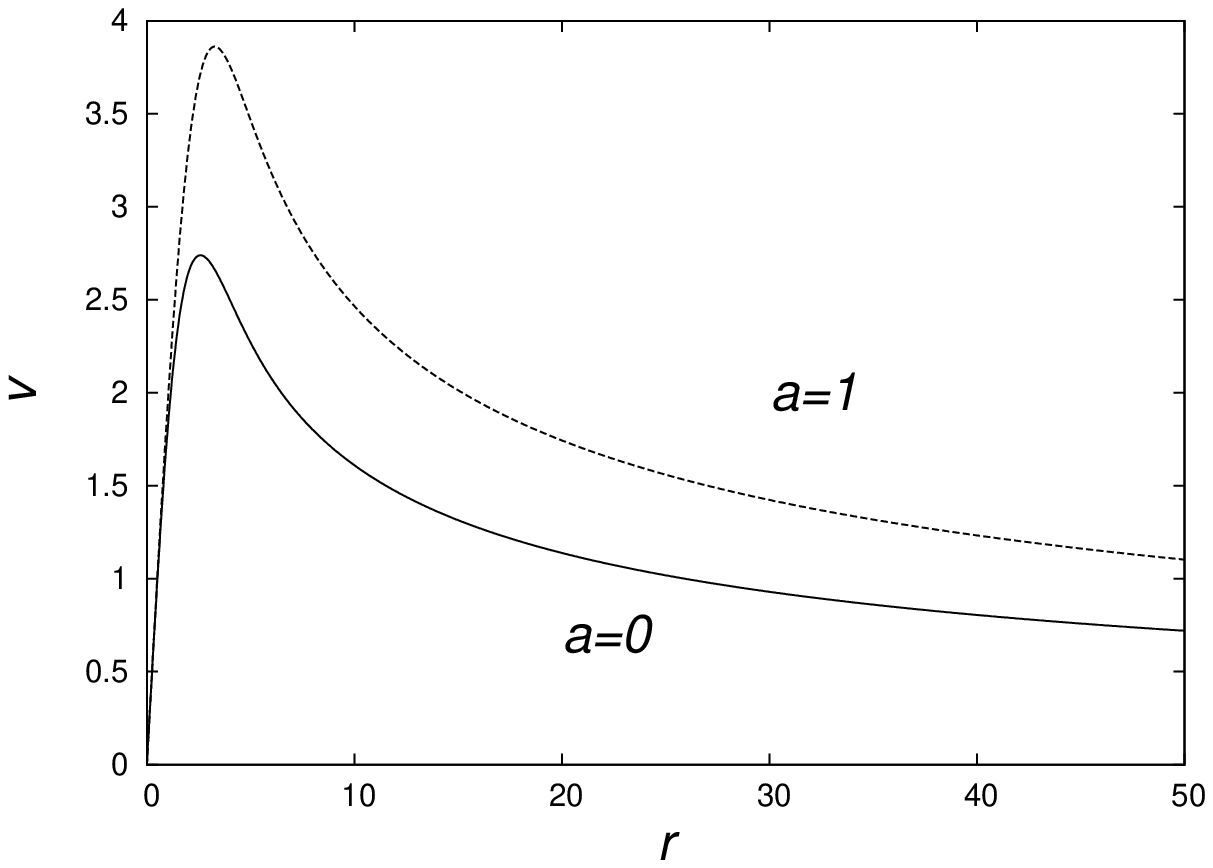}
\caption{\label{fig:equilibrium} Properties of the ground state spherically symmetric configuration for the central wave function value $\psi(0)=1$. In the top panel we show the density  of the BEC and the effect of the self-interaction term. In the bottom panel we show the effects on the rotation curve with and without self-interaction.}
\end{figure}

As example solution of (\ref{eq:eigenvalue}), we show in Fig. \ref{fig:equilibrium} the density profile and the rotation curve corresponding to the case $\psi(0)=1$ for $a=0$ and $a=1$. The introduction of a positive self-interaction allows more massive and wider BEC dark matter distributions as shown in the Figure. One of the reasons to introduce a self-interaction term, is that it might imply better RCs, however, as can also be seen in Fig. \ref{fig:equilibrium} the rotation curves with or without self-interaction show a pretty similar shape. Furthermore, the fact that self-interaction produces wider configurations does not mean that the density of the BEC is more disperse in space, instead it is more compact. Considering the examples in the figure, we have that for the case of $a=0$ the mass and compactness are $M= 25.91$, $M/r_{95}= 6.61$, whereas for $a=1$, these quantities are $M= 60.79$ and $M/r_{95}= 13.94$. Here $r_{95}$ is the radius containing 95\% of the total mass of the configuration. The failure at improving RCs when adding self-interaction, is one of the motivations to introduce some rotation, because rotation actually disperses away some of the density concentrated in the center, and the reason why we set $a=0$ from now on. 


{\it Addition of rotation.} Our code is three dimensional, and the eigenvalue problem we solve to construct spherically symmetric configurations (\ref{eq:eigenvalue}) uses spherical coordinates. Thus, in order to set initial data we interpolate the solution $\psi(r)$ of (\ref{eq:eigenvalue}) into our 3D numerical grid and obtain $\Psi(x,y,z,t=0)$. Once this has been done, we apply a rotation by redefining the wave function $\Psi = e^{-i {\bf L}\cdot {\bf \hat{n}} \theta}\Psi$. In our particular case of rotation around the $z$-axis we use $\theta = \arctan(y/x)$ with ${\bf L} = L_z \hat{\bf z}$. We  parametrize the rotation by choosing $L_z = x p_y - yp_x$ to be a constant.


\subsection{Simulations in code units}

\begin{figure}
\includegraphics[width=4.25cm]{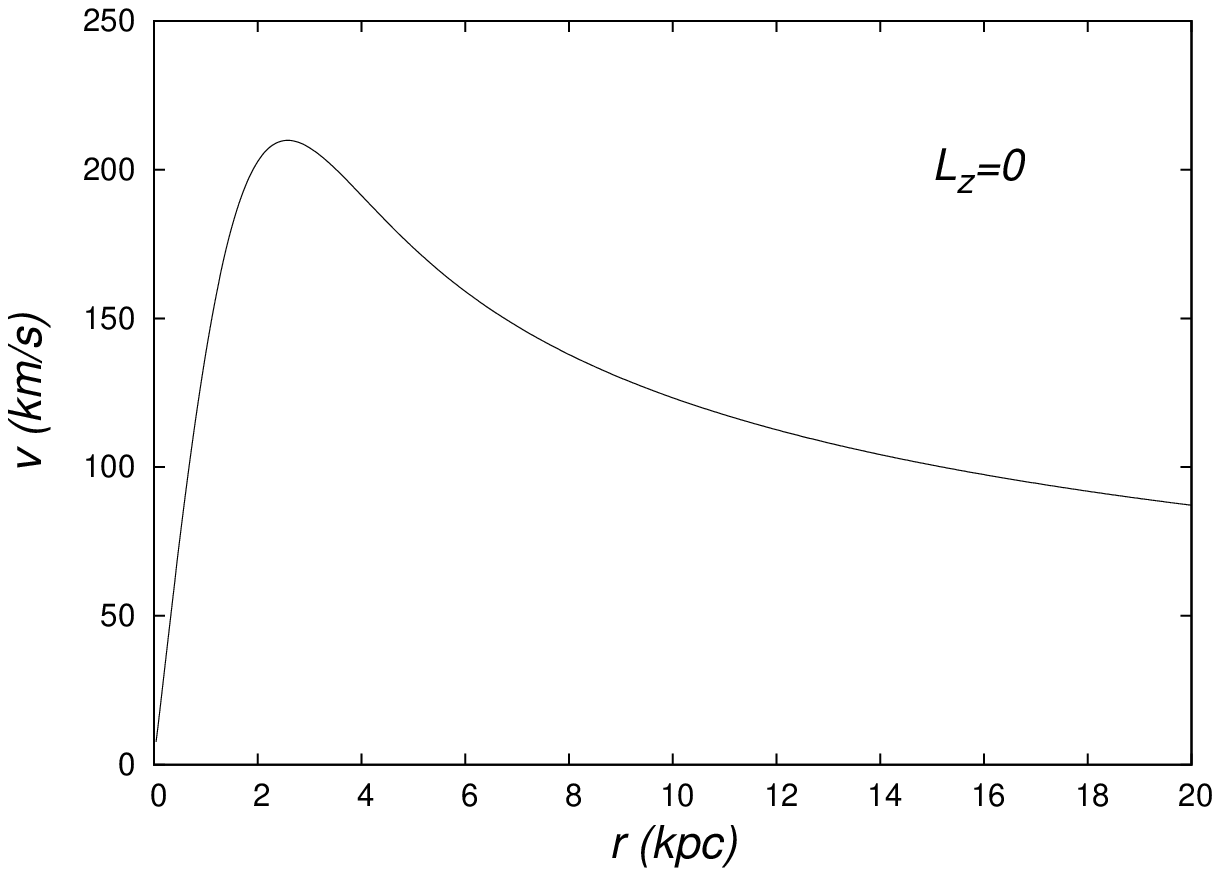}
\includegraphics[width=4.25cm]{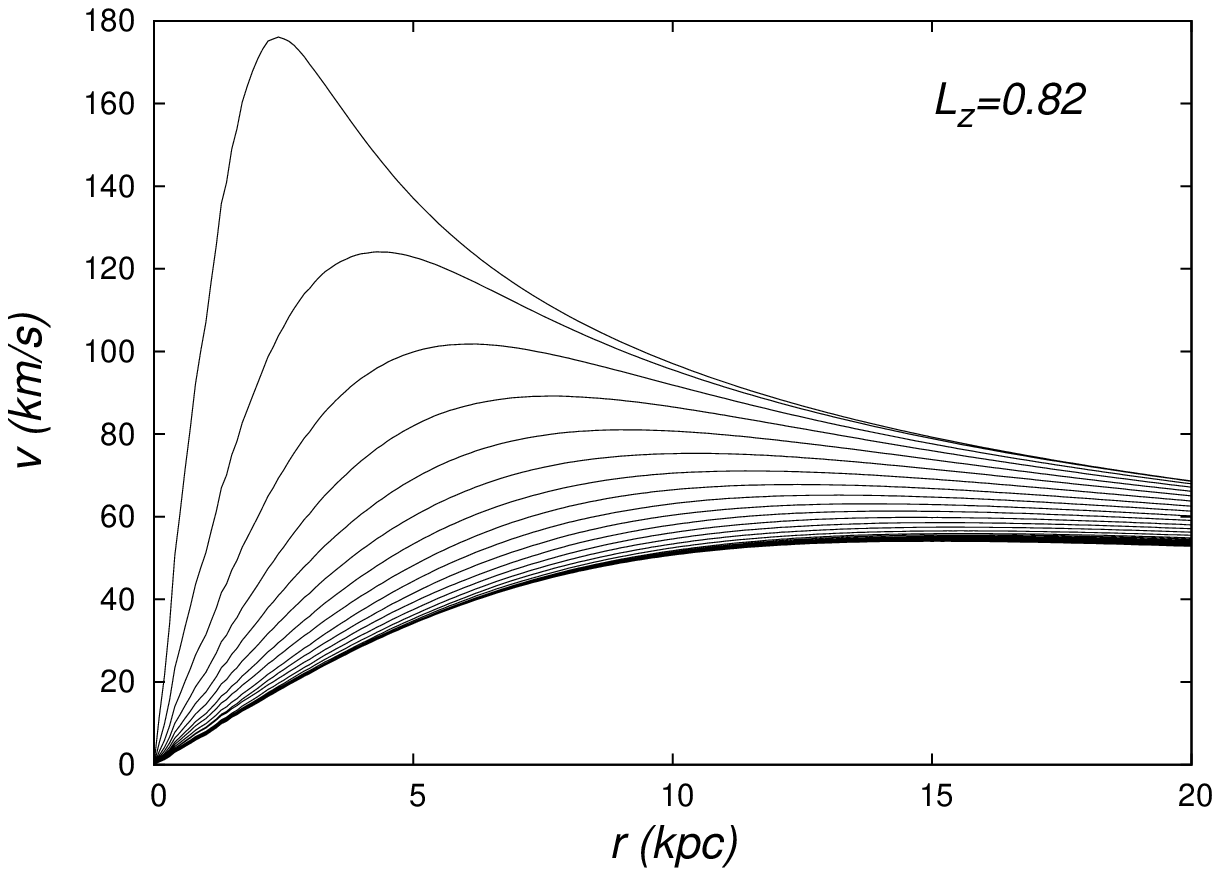}
\includegraphics[width=4.25cm]{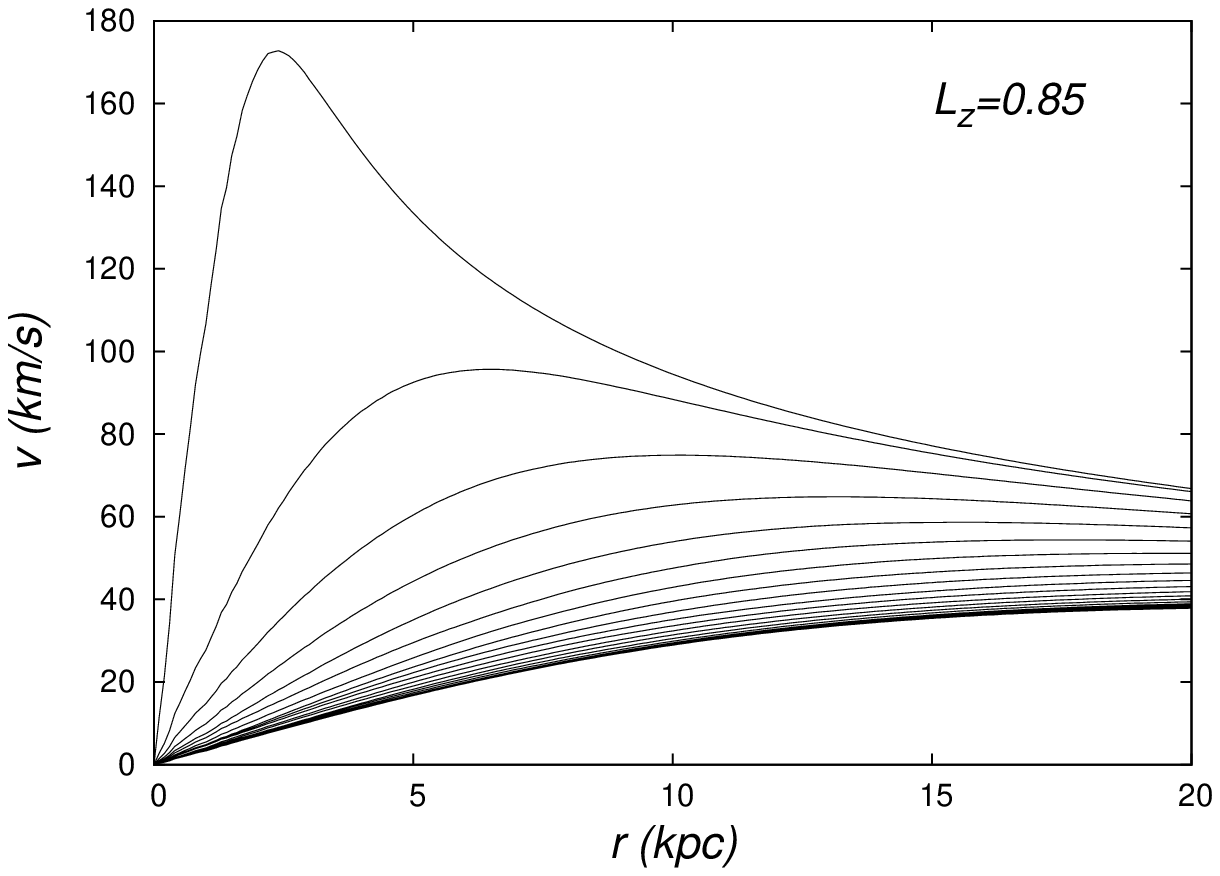}
\includegraphics[width=4.25cm]{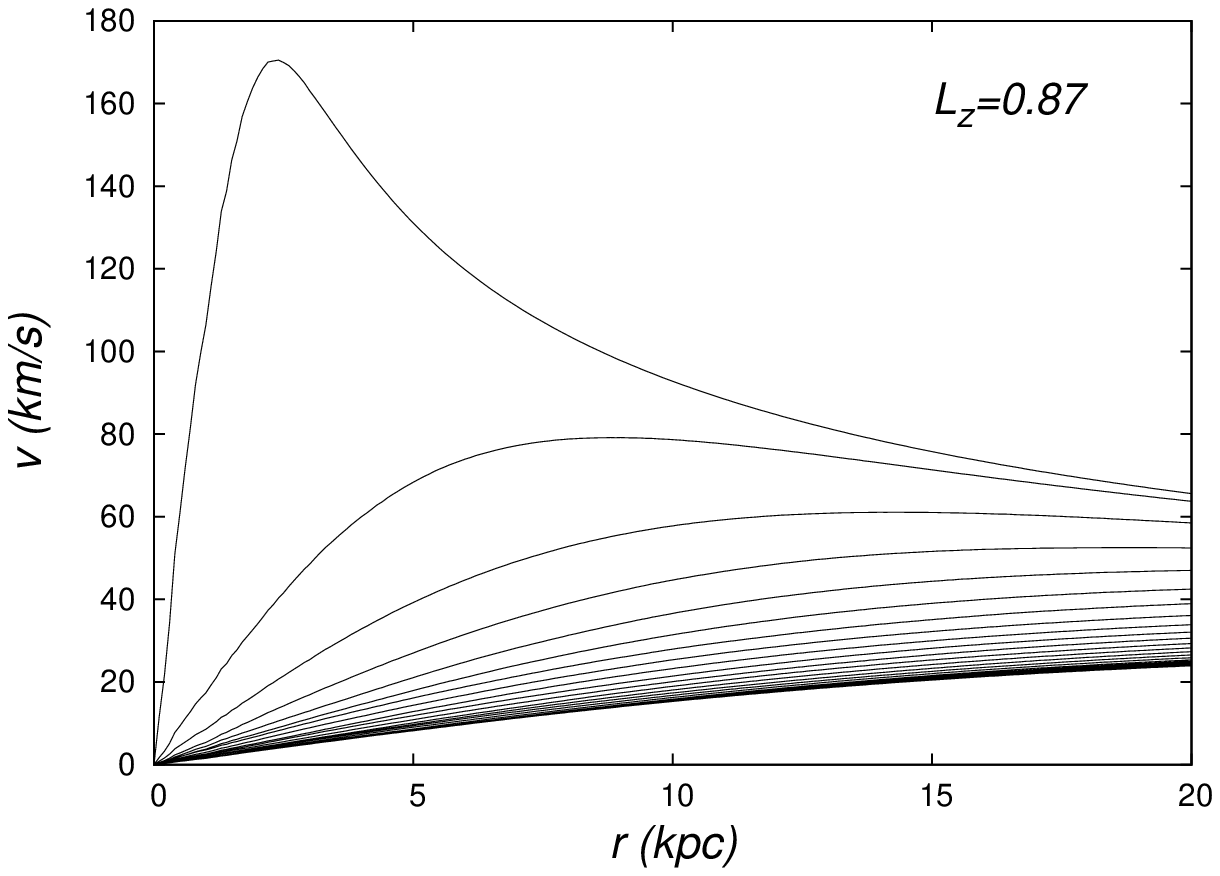}
\caption{\label{fig:long-lived} Snapshots at various times of the RCs for four values of $L_z$, which stabilize around a given profile. The time direction of the snapshots is from top to bottom in all the cases.}
\end{figure}

The criterion to consider a configuration as long-lived halo is that the rotation curve remains nearly time-independent after a transient time. We show in Fig. \ref{fig:long-lived} the RC evolution of four configurations for $\lambda=1$ and $L_z=0$ corresponding to formal equilibrium and three other values $L_z=0.82,0.85,0.87$. In the case $L_z=0$ the system remains time independent as expected, because it is a spherically symmetric equilibrium configuration; in the other three cases the RCs start high and while the density disperses away the RC flattens and starts becoming time independent. It can be seen how the snapshots start packing around a nearly time-independent profile. Such long-lived profile is the one we choose to fit the observed galactic RCs in this paper. The behavior for other values $0.82<L_z<0.87$ should show a similar behavior bounded by these two values, however the use of the three chosen values suffice to illustrate the capability of the rotating BEC halo model to fit RCs.

Additionally, we made sure that the total energy of the system is negative and the gravitational potential depth approaches an asymptotic long-living behavior in the scale of Gyr.

\subsection{From code units to physical units}

The code and physical units are related once the scale invariance parameter $\lambda$ is fixed, which we do by choosing the spatial coordinate units. Specifically, since $\lambda = \frac{\hbar}{mc}\frac{x}{\tilde{x}}$ (or equivalently the spherical coordinate $r$), if physical measurements of space $\tilde{x},\tilde{y},\tilde{z}$ are given in kpc, we choose the code coordinates $x,y,z$ to represent kpc as well. Thus it suffices to write the factor $\frac{\hbar}{mc}$ in kpc. For $m=10^{-23}$eV$/c^2$ its value is $\lambda := \lambda_0 = \frac{\hbar}{mc}[kpc]\frac{x}{\tilde{x}[kpc]}=0.0006389$.

However, the results obtained with the code units are valid for any value of $\lambda$, and in fact we parametrize the value of $\lambda=\alpha\lambda_0$ with $\alpha$ a regulation parameter that shifts the relation between code and physical units in such a way that if $\alpha>1$ a code unit of length  represents more than a kpc, whereas $\alpha <1$ does the opposite.

Once the value of $\lambda$ is fixed, we show how the relevant quantities translate from the code units into physical units.


{\it Rotation Curve velocity.} We start by writing (\ref{eq:rc}) in physical units and substitute the scaling transformations for tilde and hat variables

\begin{equation}
\tilde{v}^2 = \frac{2G}{|\tilde{x}_d|} \int |\tilde{\Psi}|^2 d\tilde{x} d\tilde{y} d\tilde{z} 
= \frac{\lambda^2 c^2}{2\pi} \frac{2 G}{|x_d|} \int |\Psi|^2 dxdydz,\nonumber
\end{equation}

\noindent and then, writing $c$ in units of $\mathrm{km/s}$ and $\lambda = \lambda_0 \alpha = 0.0006389 \alpha$, the velocity $\tilde{v}$ in $\mathrm{km/s}$ in terms of the velocity in code units $v$ relate through

\begin{equation}
\tilde{v} = \lambda \frac{c[\mathrm{km/s}]}{\sqrt{2\pi}}v = \alpha \lambda_0  \frac{c[\mathrm{km/s}]}{\sqrt{2\pi}}v \nonumber\\
= 76.597 \alpha v [\mathrm{km/s}].
\end{equation}

{\it Length.} Even though we already fixed the length units we want to specify the transformation $\tilde{x}=\frac{x}{\lambda} = \frac{x}{\lambda_0 \alpha}$. Coordinates $y,z$ scale identically.

\begin{figure}
\includegraphics[width=4.25cm]{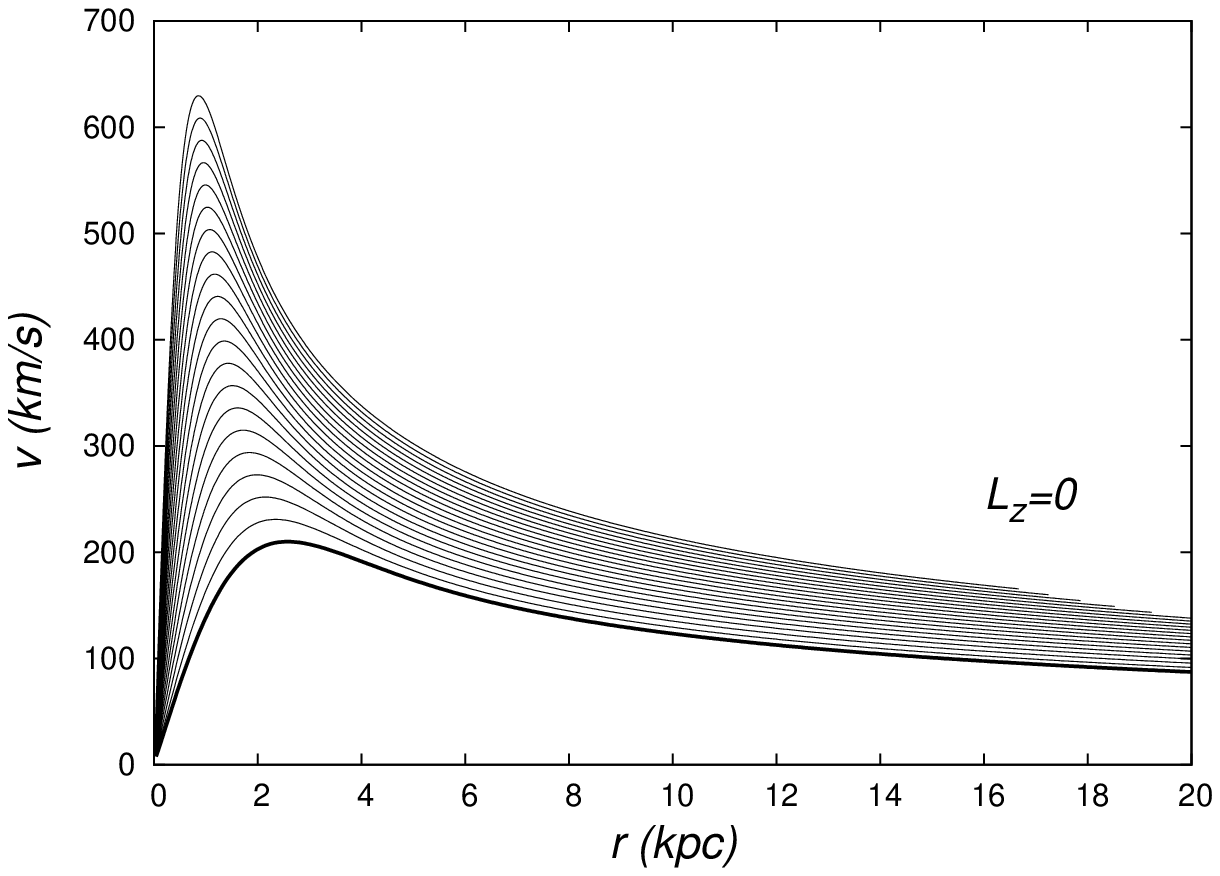}
\includegraphics[width=4.25cm]{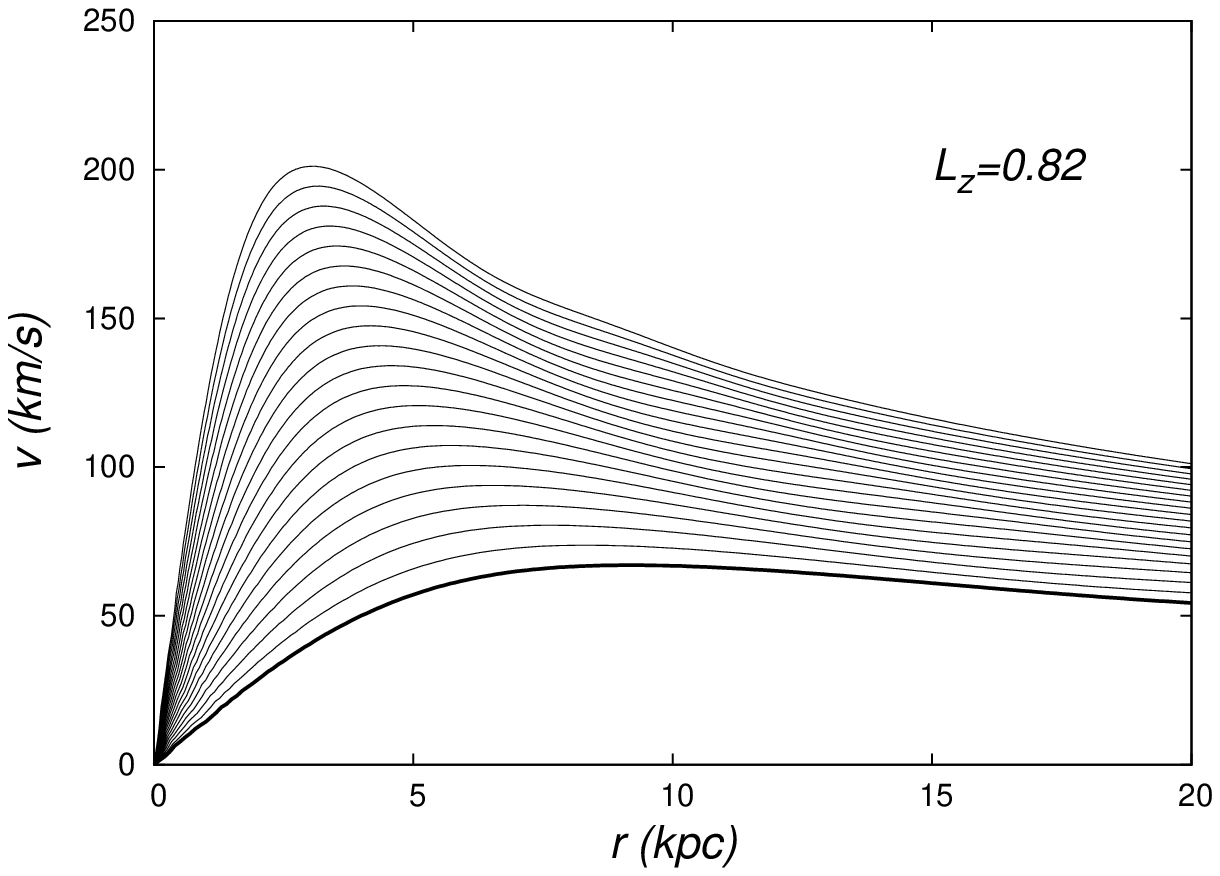}
\includegraphics[width=4.25cm]{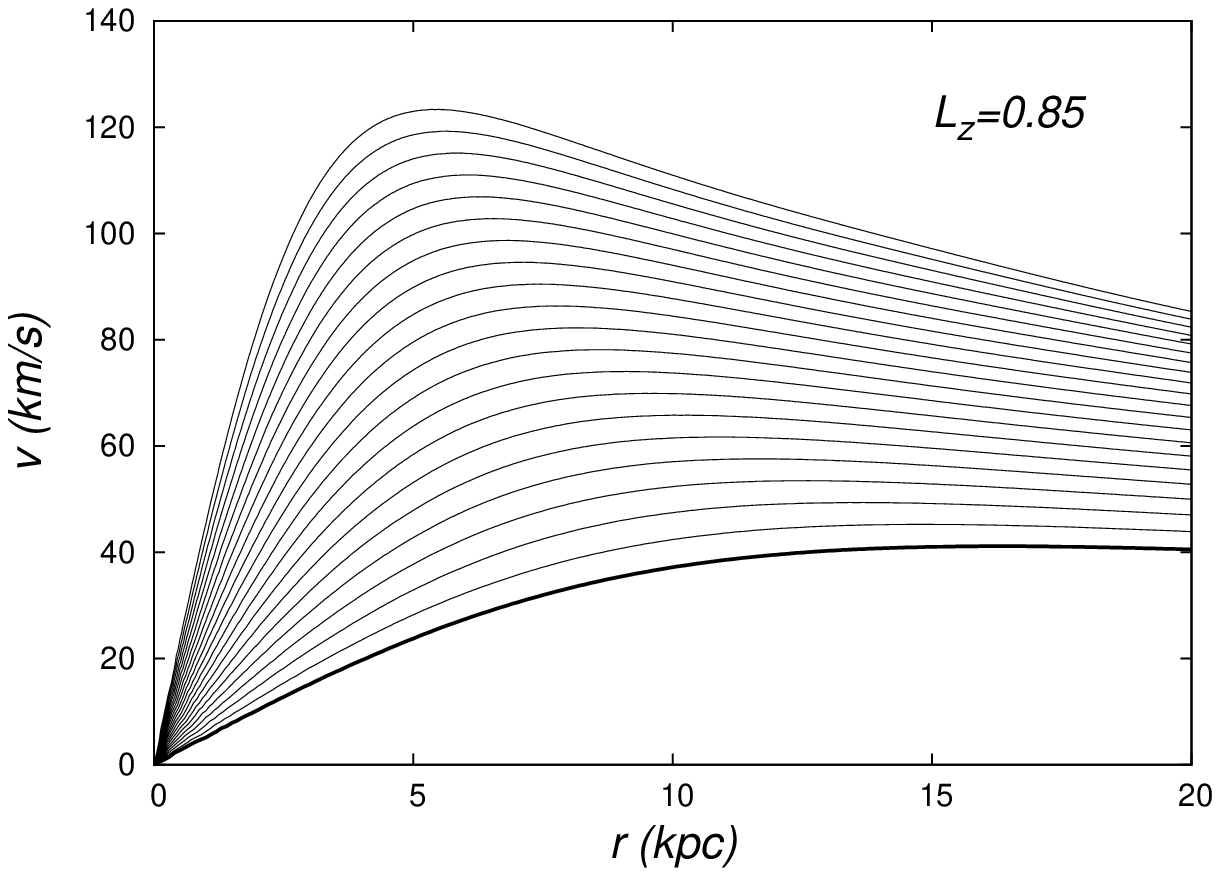}
\includegraphics[width=4.25cm]{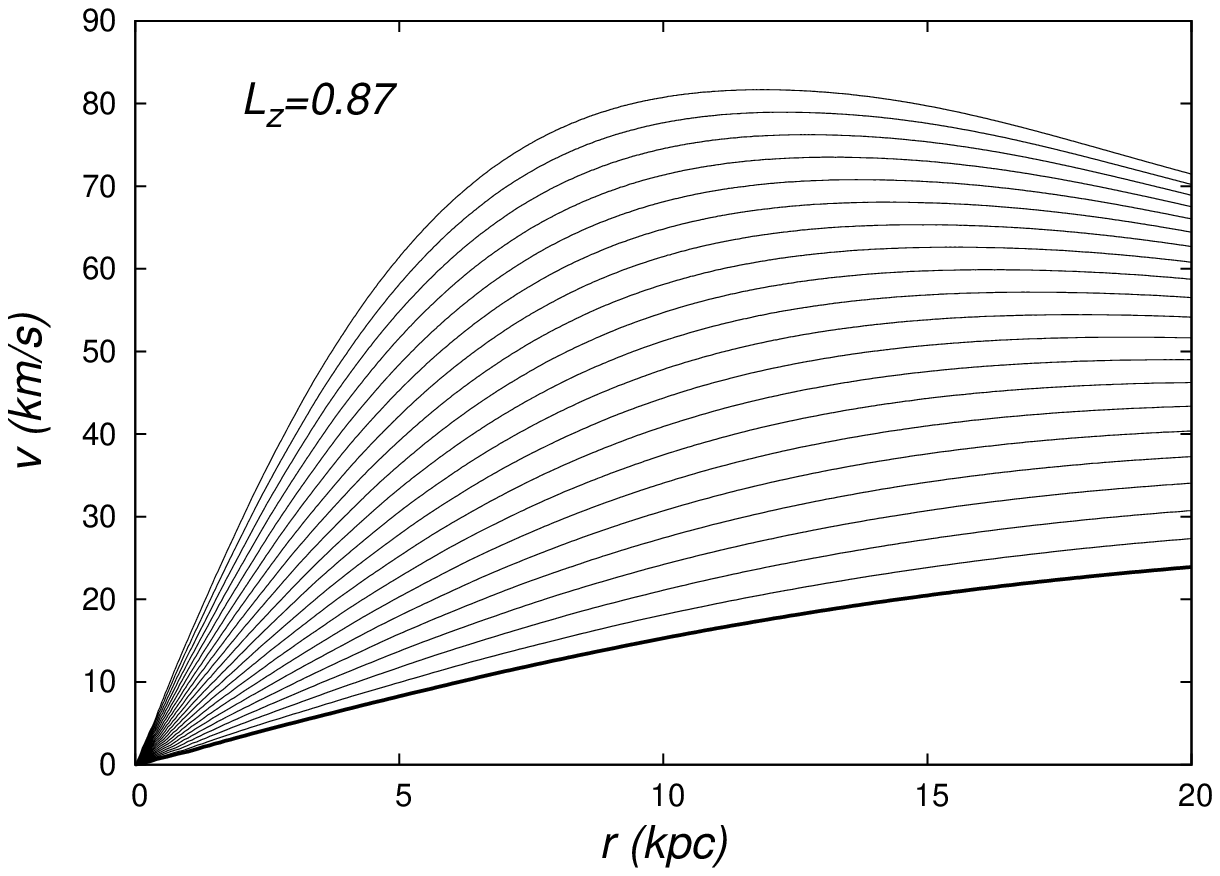}
\caption{\label{fig:RC_Lz} Variety of RCs that can be generated for different values of $L_z$. Each curve of each plot corresponds to a different value of $\alpha$. We show the curves generated for $\alpha \ge 1$ only, and the bold line corresponds to $\alpha=1$.}
\end{figure}

The scaling of velocity and length tells us about the bundle of RCs that can be constructed. By choosing $\alpha>1$ the RC in physical units shows a higher rotation curve, but at the same time it shortens the size of a halo and conversely for $0<\alpha<1$. In Fig. \ref{fig:RC_Lz} we show the bundle of different RCs produced by different values of $\alpha$ and the four chosen values of $L_z$. The larger the value of $L_z$ the more disperse the configuration is and smaller the velocity of test particles.

With these specifications, the simulations we used to construct the long-lived configurations where carried out in a domain of size $[-80\mathrm{kpc},80\mathrm{kpc}]^3$ for $\alpha=1$ and four refinement levels with resolutions 
$1.25\mathrm{kpc}$ covering the domain $[-80\mathrm{kpc},80\mathrm{kpc}]^3$,
$0.625\mathrm{kpc}$ covering the domain $[-40\mathrm{kpc},40\mathrm{kpc}]^3$,
$0.3125\mathrm{kpc}$ covering the domain $[-20\mathrm{kpc},20\mathrm{kpc}]^3$,
$0.15625\mathrm{kpc}$ covering the domain $[-10\mathrm{kpc},10\mathrm{kpc}]^3$. This domain was sufficient for the configurations to relax and approach a long-lived state.

\section{Fitting rotation curves}
\label{sec:RCs}

As mentioned before, we use values of  $L_z=0,~0.82,~0.85,~0.87$, that we found empirically to allow long-lived configurations. For each of these values we track a value of $\lambda$ that fits the RCs data. 
The sample we use to fit the RCs of our rotating halos is a subsample of LSB galaxies without luminous components in \cite{deBlok2001}.
 
 \begin{figure*}
\includegraphics[width=8cm]{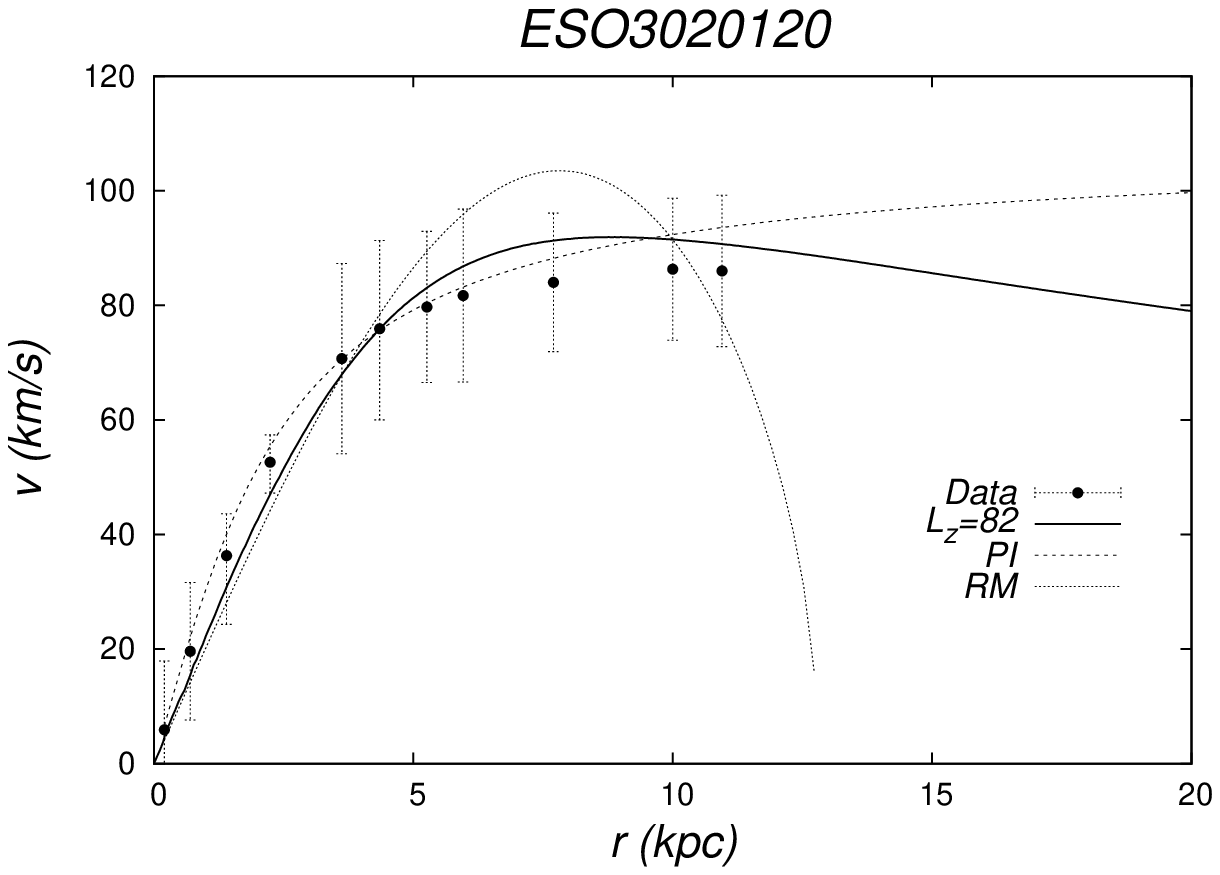}
\includegraphics[width=8cm]{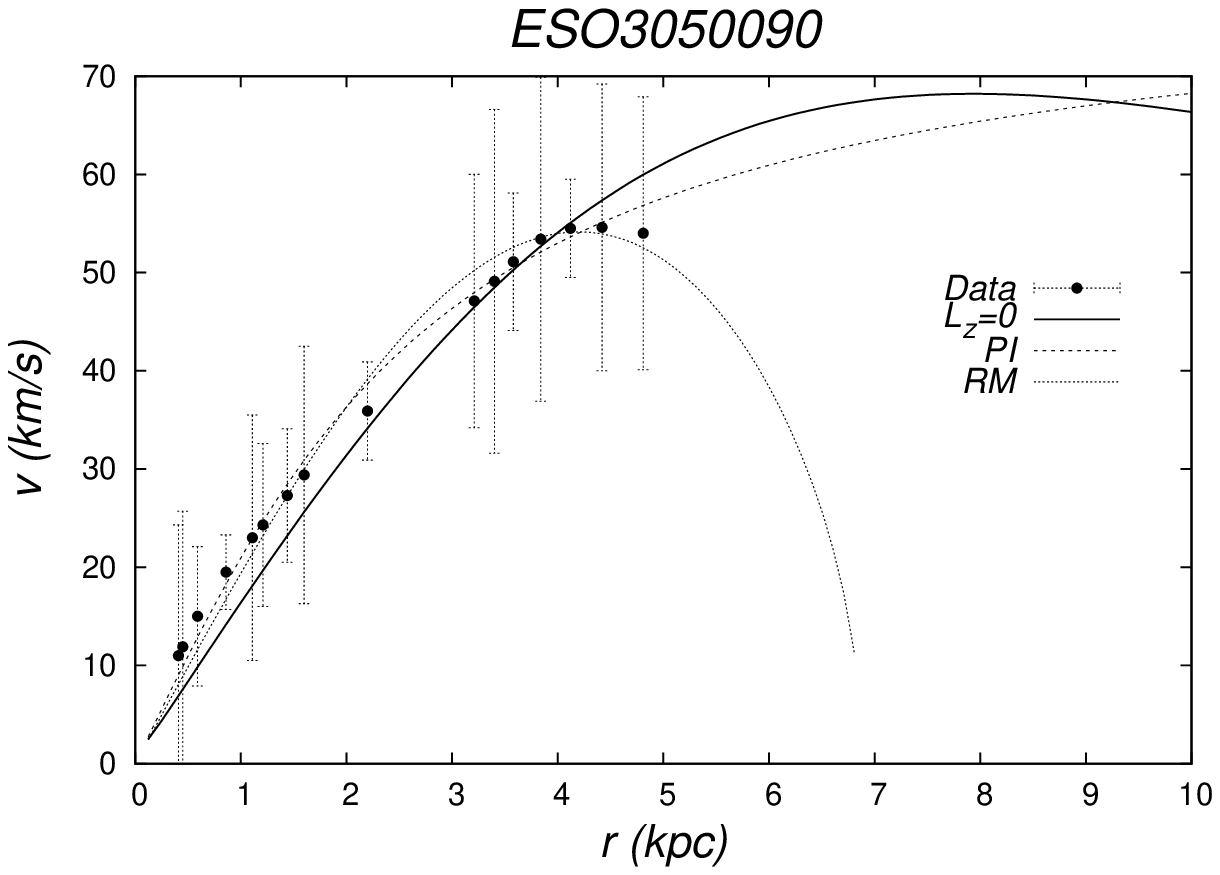}
\includegraphics[width=8cm]{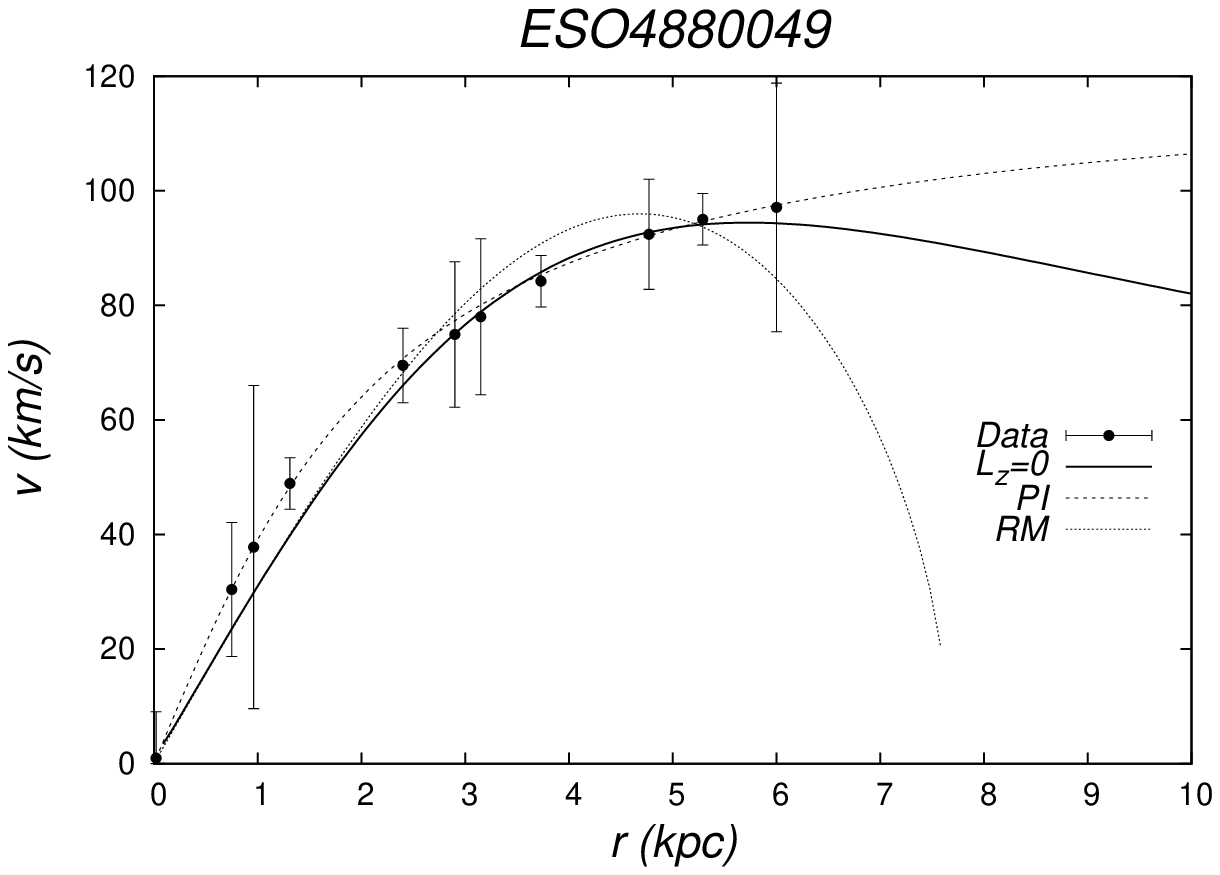}
\includegraphics[width=8cm]{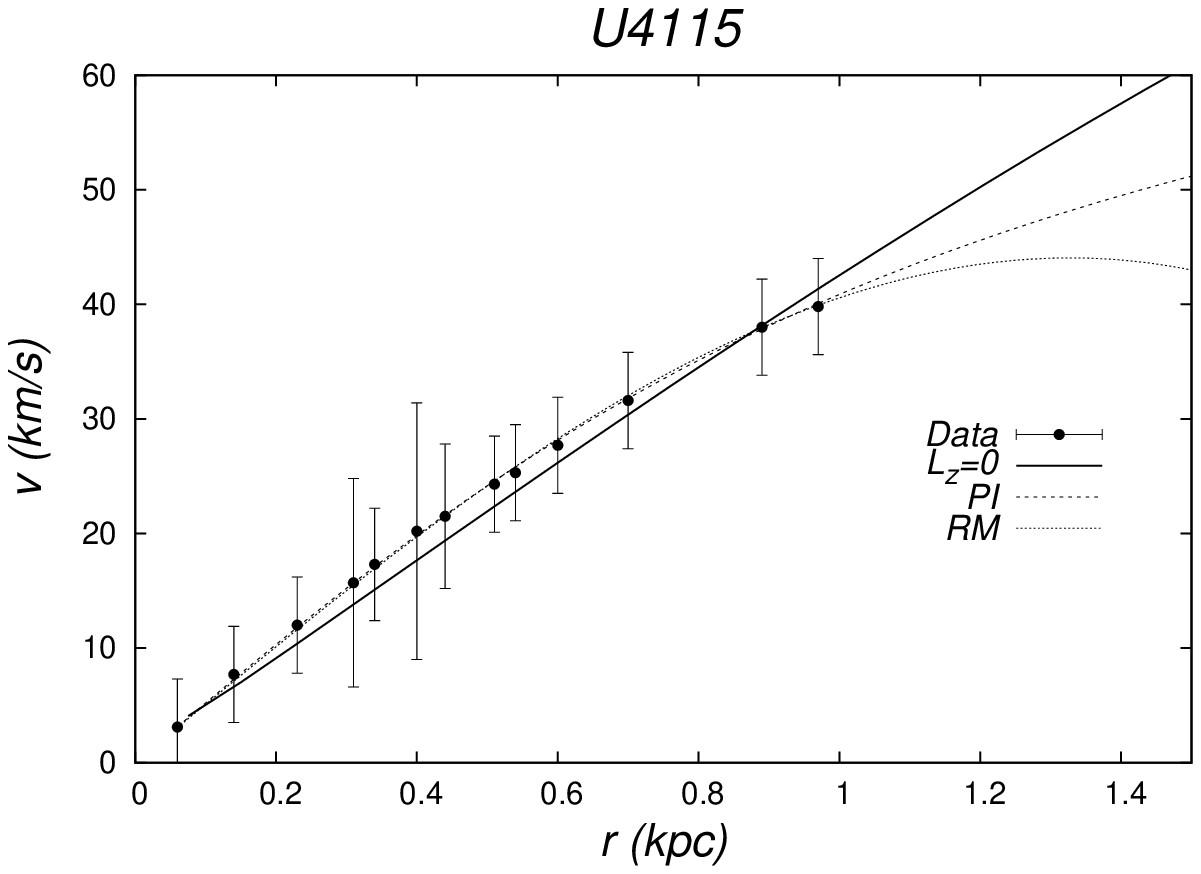}
\includegraphics[width=8cm]{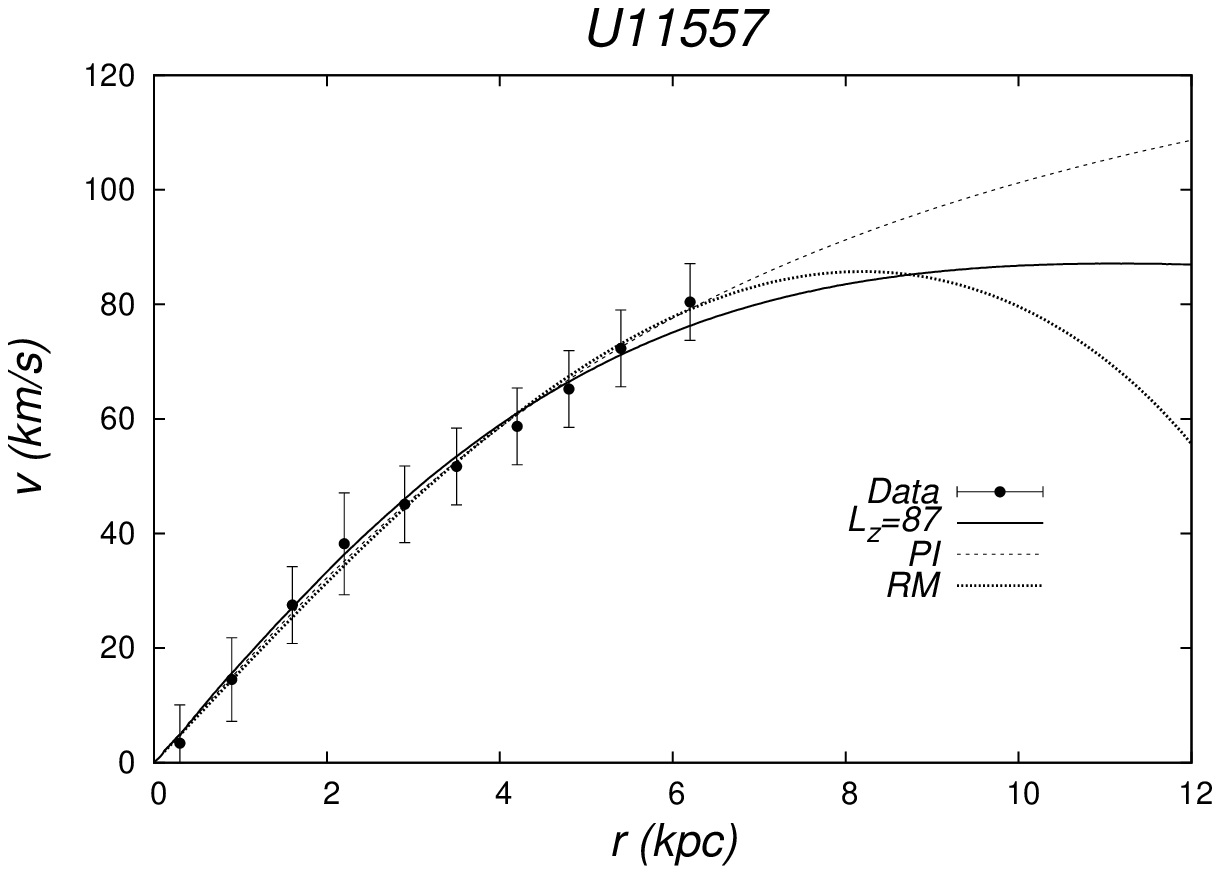}
\includegraphics[width=8cm]{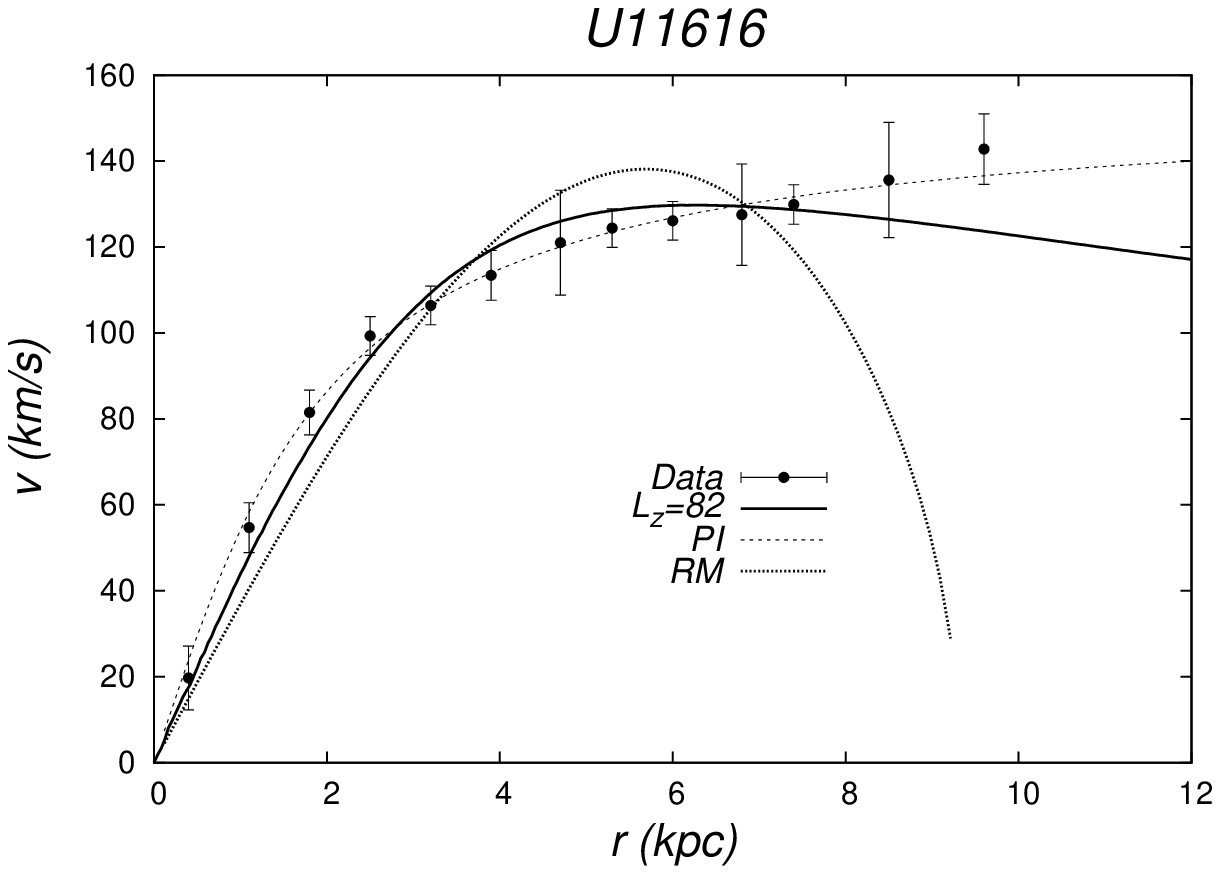}
\caption{\label{fig:RCsa0.0} RC for the galaxies ESO3020120, ESO3050090, ESO4880049, U4115, U11557 and U11616. We extend the domain in order to show the drastic fall of the curve with halos constructed in the TF limit.}
\end{figure*}

We were able to successfully fit a broad sample of LSB galaxies, as shown in Fig. \ref{fig:RCsa0.0} and the best fitting parameters are shown in Table \ref{table:a0.0}. For comparison we have also plotted the RCs due to halos made of $10^{-22}\mathrm{eV/c}^2$ bosons in the Thomas-Fermi limit constructed by Robles \& Matos (RM) in \cite{RoblesMatos2012}, which have a velocity profile given by \cite{BohemerHarko2007}

\begin{equation}
v_{RM}(r) = \sqrt{ \frac{4G \rho_0 R^2}{\pi} \left( \frac{R}{\pi r} \sin\left(\frac{\pi r}{R}\right) -\cos \left(\frac{\pi r}{R}\right)\right) }
\end{equation}

\noindent where $R$ indicates the radius at which the density of dark matter is zero. We have used the parameters $R,\rho_0$ found in \cite{RoblesMatos2012} to produce their plots. Also for comparison and control we have included the profile due to a Pseudo Isothermal profile (PI) which rotation curve profile is given by \cite{Broeils} and later used to directly fit observations \cite{deBlok2001}

\begin{equation}
v_{PI}(r) = \sqrt{ 4\pi G \rho_0 R_{PI}^{2} \left( 1- \frac{R_{PI}}{r}\arctan\left(\frac{r}{R_{PI}}\right)\right)}.
\end{equation}

\noindent We use the best fitting parameters $\rho_0, R_{PI}$ found in \cite{RoblesMatos2012}. 
We have extended the spatial domain on purpose, so that future experimental points may decide between the models constructed assuming the Thomas-Fermi limit and our more general model.

A clear difference between our halo model of BEC dark matter and that assuming the Thomas-Fermi limit, is that in the later case there is a cut off of the density at a finite radius that makes the rotation curve to drop drastically. Clear cases are the galaxies ESO3020120, ESO4880049 and U11616 in Fig. \ref{fig:RCsa0.0}, where the rotation curve of the RM model fits only in the inner parts, but clearly drops further out. In these cases our rotating model performs much better and shows a trend more similar to the PI control profile.

\begin{table}
\begin{tabular}{ccc}\hline
Galaxy 		& Best-$L_z$ 	& $\alpha$\\\hline
ESO3020120	& 0.82 	& 1.7\\
ESO3050090	& 0		& 0.325\\
ESO4880049	& 0		& 0.45\\
U4115 		& 0		& 0.53\\
U11557 		& 0.87	& 3.2\\
U11616 		& 0.82	& 3.5\\\hline
\end{tabular}
\caption{\label{table:a0.0} Best values of $L_z$ and $\alpha$ that fit the RCs.}
\end{table}


\section{Discussion and Conclusions} 
\label{sec:conclusions}

Given the diversity of RCs in LSB galaxies, we explored the chances that different values of $L_z$ could also explain such diversity. We have shown that fixing $m$ and $a$ (in our case $a=0$), a bundle of RCs with different profiles can be constructed for each value of $L_z$.  

A more general analysis of the parameter space includes the variation $m$ and $a$, which would imply a three dimensional parameter space. This situation rather defines an inverse problem. That is, an analysis looking toward the BEC dark matter model will require data analysis of a universal sample of galaxies in order to fix a single value of $m$ and $a$, so that the same bosons are the same dark matter in every halo, a mistake usually overseen in models in the Thomas-Fermi limit, where different values of $a$ are found for different galaxies, which is somehow inconsistent (e. g. \cite{BohemerHarko2007,RoblesMatos2012}).

Our results show that rotating BEC dark matter halos, and non-rotating ground state equilibrium configurations are an option worth to study in more general samples of galaxies, in which luminous matter has a more considerable contribution. This however will require the development of a code that in the minimal case solves the GPP system coupled to Euler equations to describe the luminous matter.

Finally, even though our working hypotheses of an ultralight boson and zero self-interaction are similar to the only structure formation analysis so far in \cite{Schive}, it would be also interesting to consider the recent restrictions found, imposed by BBN on $m$ and $a$ \cite{ShapiroCosmoConstraints,ShapiroModPhys}.


\section*{Acknowledgments}

We appreciate the comments from the anonymous referee.
This research is partly supported by grants: 
CIC-UMSNH-4.9 and 
CONACyT 106466.
F.D.L-C gratefully acknowledges a postdoctoral DGAPA-UNAM grant.
F.S.G. acknowledges support from the CONACyT program for sabbatical visits in foreign countries and the kind hospitality of the Physics and Astronomy Department at UBC, where part of this paper was developed.


\end{document}